\documentclass[12pt]{article}

\def\ii{\'{\i}}

\usepackage{epsfig}

\def\arrowright{\smash{\mathop{\longrightarrow}\limits_{\raise3pt\hbox{$\eta \to 0$}}}}

\def\tozero{\smash{\mathop{\longrightarrow 0}\limits_{\raise3pt\hbox{$\eta \to 0 , \eta \to \infty$}}}}

\def\infinito{\smash{\mathop{\longrightarrow}\limits_{\raise3pt\hbox{$\eta \to \infty$}}}}

\def\Ninfinito{\smash{\mathop{\longrightarrow}\limits_{\raise3pt\hbox{$N_A \to \infty$}}}}

\begin{document}

\title{Multiplicity fluctuations and percolation of strings in hadron-hadron and nucleus-nucleus collisions}%
\author{Pedro Brogueira\footnote{Department of Physics, Instituto Superior T\'{e}cnico, 1049-001 Lisboa, Portugal} and  
Jorge Dias de Deus$^*,$\footnote{CENTRA, Centro Multidisciplinar de Astrof\ii sica}}
\maketitle

\begin{abstract}
We argue that recent NA49 results on multiparticle distributions and fluctuations, as a function of the number of participant nucleons, suggest that percolation plays an important role in particle production at high densities. 
\end{abstract}


\medskip

Recentely, the NA49 collaboration has presented results, from the CERN/SPS at 158 A GeV, on multiplicity fluctuations or, to be more precise, on $V(n)/\langle n\rangle$, 

\begin{equation}
V(n)/\langle n\rangle \equiv \frac{\langle n^2\rangle-\langle n\rangle^2}{\langle n\rangle},
\end{equation}
as a function of the number $N_{part.}$ of participant nucleons, from $pp$ to $PbPb$ collisions [1].

These data are very interesting for several reasons:

1) They show evidence for universal behaviour: the experimental points in the plot $V(n)/\langle n\rangle$ versus $N_{part.}$ fall into a unique curve (see Fig.1).

\begin{figure}[t]
\begin{center}
\includegraphics[width=10cm]{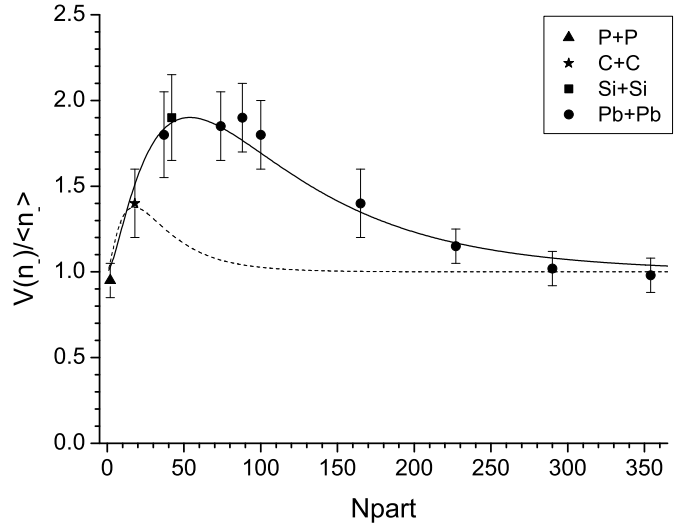}      
\end{center}
\caption{Variance over average multiplicity, for negative particle production, as a function of the number of participants. The curves are from (16) (full curve, for NA49 data, dashed curve for RHIC, $\sqrt{s}=200$GeV).}
\end{figure}

2) The physics in the small $N_{part.}$ limit $(pp,N_{part.}\to 2)$ and in the large $N_{part.}$ limit $(PbPb, N_{part.}\to 2A_{PbPb})$ seems to be quite the same, as in both cases the quantity (1) approaches 1. The fluctuations are larger in the intermediate $N_{part.}$ region (see Fig.1).

3) The (negative) particle distribution, in the low density and in the high density limits, is in fact a Poisson distribution (see Fig.2), the distribution being wider than Poisson in the intermediate $N_{part.}$ region.

\smallskip

In the framework of the string model with percolation [2], these results are natural. On one hand, percolation is a universal geometrical phenomenon, the properties depending essentially on the space dimension (dimension 2, impact parameter plane, in our case), and being controlled by the transverse density variable $\eta$,
\begin{figure}
\begin{center}
\includegraphics[width=8cm]{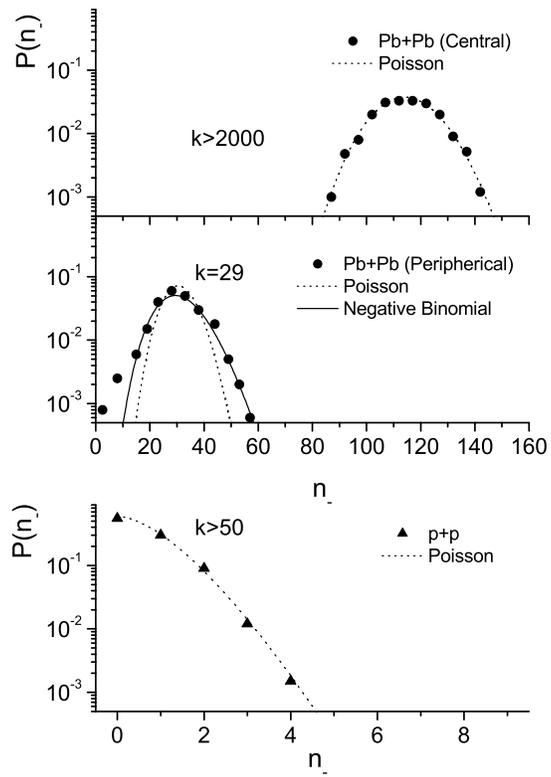} 
\end{center}   
\caption{Multiplicity Distributions, $P(n_-)$, as a function of $n_-$. The curves are Poisson (dashed lines) and Negative Binomial (full line).} 
\end{figure}

\begin{equation}
\eta \equiv \left( {r\over R}\right)^2 \bar N_s \ ,
\end{equation}
where $r$ is the transverse radius of the string $(r\simeq 0.2 fm)$, $R$ the radius of the interaction area, and $\bar N_s$ the average number of strings. The quantity $(R/r)^2$ is nothing but the interaction area in units of the string transverse area. As $R$ and $\bar N_s$ are functions of the number $N_{part.}$ of participants, $N_{part.}$, similarly to $\eta$, becomes, at a given energy,a universal variable.

On the other hand, in percolation [3], what matters is the fluctuation in the size of the clusters of strings: one starts, at low density (small $N_{part.}$), from a situation where strings are isolated, at intermediate density one finds clusters of different sizes, and one ends up, at high density, above the percolation threshold, with a single large cluster. In both, low and high, density limits, for a fixed number $\bar N_s$ of strings, fluctuations in cluster size vanish (see Fig.3). In the simplest string model the particle distribution is Poisson (as observed in $e^+e^-$ and $pp$ at low energy) and $V/\langle n\rangle \to 1$ in both, low and high, density limits (as seen in Figs.1 and 2).

In hadron-hadron and nucleus-nucleus collisions, during the collision strings are produced along the collision axis, and these strings may overlap and form clusters of different sizes. In the spirit of percolation theory, we shall assume that fluctuations in the number $N$ of strings per cluster dominate over all the other fluctuations.

Following [4] we write for the multiplicity distributions at a given value of $\eta$ (or $N_{part.}$)
\begin{equation}
P(n)=\int W(X) p(n,X\bar n)dx \ ,
\end {equation}
where the integral is over the cluster distribution $W(X)$, and $p(n,X\bar n)$ is a convolution of Poisson distributions, $\bar n$ being the single string average multiplicity. If $W(X)$ is a gamma function, $P(n)$ becomes a negative binomial (NB) distribution. From (3) we obtain
\begin{equation}
\langle n\rangle = \langle X\rangle \bar n \ ,
\end{equation}
and
\begin{equation}
1/k \equiv {\langle n^2\rangle -\langle n\rangle^2 \over \langle n\rangle^2} - {1\over \langle n\rangle } = {\langle X^2\rangle - \langle X\rangle^2 \over \langle X \rangle^2} \ ,
\end{equation}
$k$ being the NB parameter. As, see [5],
\begin{equation}
X=\bar N_c N \ ,
\end{equation}
where $\bar N_c$ is the average number of clusters and $N$ the number of strings in a given cluster, and the sum-rule,
\begin{equation}
\bar N_s =\bar N_c \langle N\rangle 
\end{equation}
is valid, we have, instead of (4) and (5),
\begin{figure}
\includegraphics[width=6cm,angle=270]{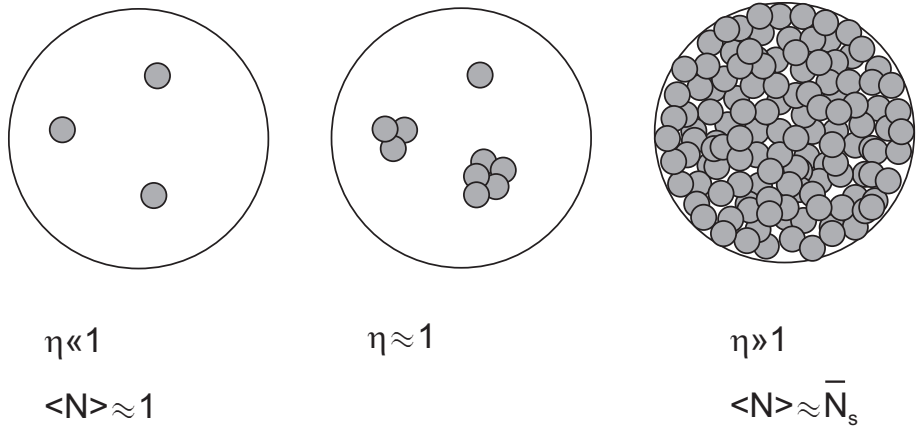}
\caption{Impact parameter percolation. For small densities $(\eta \ll 1)$ and for large densities $(\eta \gg 1)$ there are no strong fluctuations in the number $N$ of strings per cluster. For intermediate densities $(\eta \simeq 1)$ $N$-fluctuations are large.}
\end{figure}
\begin{equation}
\langle n\rangle = \bar N_s \bar n \ ,
\end{equation}
and
\begin{equation}
1/k = {\langle N^2\rangle -\langle N\rangle^2 \over \langle N\rangle^2} \ .
\end{equation} 

As the NA49 experiment is at relatively low energy ($\sqrt{s} \simeq 20$GeV) we shall have contributions from valence strings $(V)$ and from sea strings (5). We thus write, 
\begin{equation}
\bar n = {\bar N^V_s \over \bar N_s} \bar n^V + {\bar N_s^S \over \bar N_s} \bar n^S \ ,
\end{equation} 
with
\begin{equation}
\bar N_s = \bar N^V_s + \bar N_s^S \ ,
\end{equation}
with, as valence strings are longer, $\bar n^V > \bar n^S$. We then have, from (8) and (10),
\begin{equation}
\langle n\rangle = \bar N_s^V \bar n^V + \bar N_s^S \bar n^S \ ,
\end{equation}
and, from (1) and (9),
\begin{equation}
{V (n)\over \langle n\rangle} = {\langle n\rangle \over k} +1 \ .
\end{equation}

>From the percolation arguments given above, $\langle N^2\rangle - \langle N\rangle^2 \to 0$ as $\eta \to 0$ and $\eta \to \infty$, which implies $k(\eta) \to \infty$ (Poisson distribution) in the same limits. As $\langle n\rangle$, similarly to $\bar N_s$, is a monotomically increasing function of $\eta$ (and of energy and of $N_{part.}$) it is clear that from (13) one expects a maximum of $V(n)/\langle n\rangle$ at same value of $\eta$ and $V(n)/\langle n\rangle \to 1$ at low and density, as seen in data.

An attempt was made in [5] to study the $k$ dependence on $\eta$ in $pp (\bar pp)$ collisions. Experimentally, at low $\eta$ (or energy) $k$ is large and decreases with increasing $\eta$. This is true in percolation and is true as well in the "coins-in-boxes" model utilized in [5]. For the fluctuation of the number $N$ of coins per box one obtains in that model
\begin{figure}
\begin{center}
\includegraphics[width=11cm]{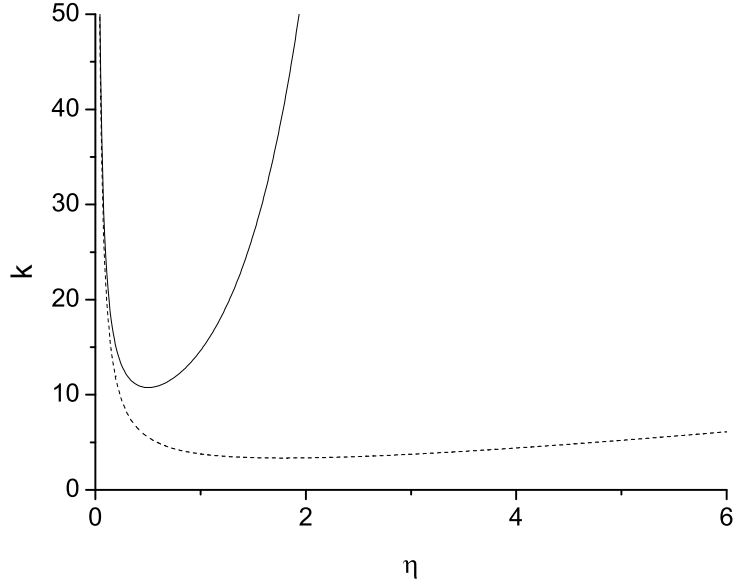}
\end{center}
\caption{The parameter $k$ as a function of $\eta$: the "coins in boxes" model (dashed curve) and the percolation model (full curve).}
\end{figure}

\begin{equation}
k(\eta ) = {\eta \over 1-(1+\eta) exp(-\eta )}
\end{equation}
and the curve (14) is shown in Fig. (4) - dashed line. At low density $\eta$ the "coins in boxes" model is similar to percolations: strings are isolated, coins sit in different boxes. At high density, as the boxes do not loose their identity, the equivalent to the appearence of a single big cluster does not occur: this means that fluctuations do not vanish at $\eta \to \infty$, as it happens with percolation.

In order to have a model satisfying the fix $\bar N_s$ percolation conditions

\begin{equation}
\langle N^2\rangle - \langle N\rangle^2 \tozero
\end{equation}
we modified the "coins in boxes" model, (14), to write
\begin{equation}
k(\eta) = b {e^{b\eta} -1\over 1-(1+b\eta)e^{-b\eta}}   \ ,
\end{equation}
such that (16) agrees with (14) at low $\eta$, but $k (\eta)$ in (16) increases much faster at high density (see full curve in Fig. (4).

Before making a comparison with NA49 data, there are two questions to be addressed:

i) {\it $F(\eta)$ factor due to random colour summation}

When strings fuse in a cluster the effective colour charge is not just the sum of the colour charges of the individual strings [6]. In pratice, the effective number $N$ of strings is reduced, [7],
\begin{equation}
N \longrightarrow F (\eta) N \ ,
\end{equation}
where 
\begin{equation}
F(\eta) = \sqrt{1-e^{-\eta}\over \eta} \ ,
\end{equation}
such that, instead of (12),
$$
\hskip 4 true cm \langle n\rangle = F(\eta) \left[ \bar N_s^V \bar n^V +\bar N_s^S \bar n^S \right] \ , \hskip 3,5 true cm (12')
$$
but (13) remains unchanged.

ii) {\it The relation between $\eta$ and $N_{part.}$}

In the definition of $\eta$, (2), what appears is not $N_{part.}$ but rather the average number $\bar N_s$ of strings and the radius $R$ of interaction. Making use of simple nuclear physics and multiple scattering arguments, one has [8]
\begin{equation}
R \simeq R_p N_A^{1/3} \ ,
\end{equation}

\begin{equation}
\bar N_s \simeq \bar N_s^p N_A^{4/3} \ , 
\end{equation}
and 
\begin{equation}
\bar N_s^V \simeq 2 N_A \ .
\end{equation}
where $R_p$ is the nucleon radius $(\simeq 1fm)$, $\bar N_S^p$ is the (energy dependent) number of strings in $pp$ collisions, at the same energy, and $N_A$ is given by 
\begin{equation}
N_A = {N_{part.}\over 2} \ .
\end{equation}

>From (2), (19), (20) and (22) we obtain for the relation between $\eta$ and $N_{part.}$,
\begin{equation}
\eta = \left( {r\over R_p}\right)^2 \bar N_s^p N_A^{2/3} = \left( {r\over R_p}\right)^2 \bar N_s^p \left( {N_{part.}\over 2}\right)^{2/3} 
\end{equation}

We turn now to the comparison with experiment:

\medskip

{\bf 1- Multiplicity $\bf \langle n\rangle$}
\smallskip

Instead of using directly (12') we shall write $\langle n\rangle_{N_A}$ as a function of $\langle n\rangle_p$, making use of (12'), (19), (20), (21), (22) and (23),
\begin{equation}
\langle n\rangle_{N_A} = {F(\eta_{N_A})\over F(\eta_p)} \langle n\rangle_p N_A^{4/3} {1+c N_A^{-1/3}\over 1+c} \ ,
\end{equation}
with
\begin{equation}
c = {2 \left( \bar n^V -\bar n^S \right) \over \bar n^S \bar N_s^p}
\end{equation}
being an adjustable parameter, $c\ge 0$, decreasing with energy.

In Fig. (5) we show (17) in comparison with NA49 data.
\begin{figure}
\begin{center}
\includegraphics[width=10cm]{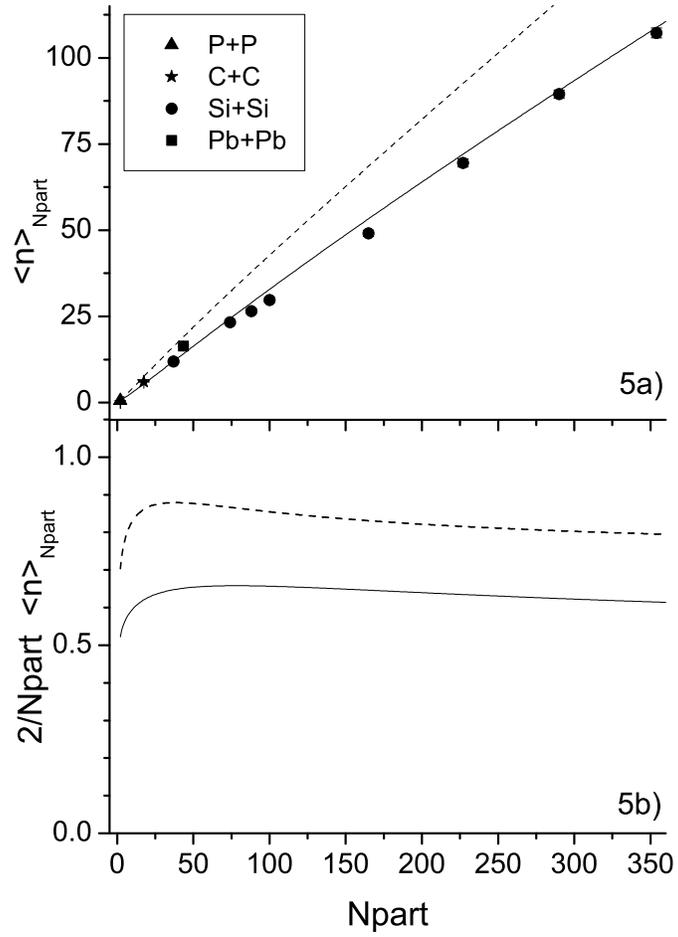}
\end{center}
\caption{a) The multiplicity as a function of $N_{part.}$ in comparison with NA49 data (full line). Prediction at RHIC, $\sqrt{s}=200$GeV (dashed curve); \ b) The ratio  $\langle n\rangle /N_A$ as function of $N_{part.}$ (full line for NA49 data and dashed line for $\sqrt{s}=200$GeV).}
\end{figure}
\medskip

{\bf 2- Fluctuations: $\bf V\langle n\rangle / \langle n\rangle$}
\smallskip

In Fig. (1) we show the comparison of the percolation model (16) with the NA49 data, and the prediction at $\sqrt{s}=200$GeV. 

The parameters used in both fits, Figs. (5) and (1) were: $b=1.68$, $c=2.3,\langle n\rangle_p =0.52$ (from data), $\bar N_s^p=3.5, \left( R_p/r\right) \simeq 5$. 

In Figs. (5) and (1) the dashed lines represent our prediction (without phase space adjustments) for RHIC ($\sqrt{s} =200$GeV), with $\bar N_s^p =7.5$ and $\langle n\rangle_p =0.70$ (see first paper in [2]).

In conclusion, we find that the recent NA49 results, regarding the multiplicity distribution dependence on the number of participating nucleons are quite consistent with the impact parameter, percolation description of hadron-hadron and nucleus-nucleus collisions at high density and high energy, as proposed in [4] and [5]. For related work see [9] and [10].

We would like to thank Elena Ferreiro, Carlos Pajares and Roberto Ugoccioni for many discussions, and to thank P. Seyboth and M. Rybcz\'ynski for information on NA49 data. This work has been done under the contract POCTI/36291/FIS/2000, Portugal.

\vfill \eject
\textbf{References:}

\begin{enumerate}
\item M. Ga\v zdzicki et al., J. Phys. G30, 5701 (2004) (nucl-ex, 0403023).
\item N. Armesto, M.A. Braun, E.G. Ferreiro and C. Pajares, Phys. Rev. Lett. 77, 3736 (1996); M. Nardi and H. Satz, Phys. Lett. B442, 14 (1998); A. Rodrigues, R. Ugoccioni and J. Dias de Deus, Phys. Lett. B458 (1999) 402.
\item D. Stauffer, Phys. Rep. 54, 2 (1979); D. Stauffer and A. Aharony, Introduction to Percolation theory, Taylor and Francis (1992).
\item J. Dias de Deus, E.G. Ferreiro, C. Pajares and R. Ugoccioni, Euro Phys. Journal, to appear, (hep-ph/0304068); C. Pajares, Acta Phys. Pol. B35, 153 (2004).
\item J. Dias de Deus,  E.G. Ferreiro, C. Pajares and R. Ugoccioni, Phys. Lett. B601 (2004) 125.
\item T.S. Biro, H.B. Nielsen and J. Knoll, Nucl. Phys. B245, 449 (1984).
\item M.A. Braun, F. Del Moral and C. Pajares, Phys. Rev. C65, 02490 (2002).
\item J. Dias de Deus and R. Ugoccioni, Phys. Lett. B491, 253 (2000), Phys. Lett. B494, 53 (2000).
\item P. Brogueira and J. Dias de Deus, Acta Phys. Polon. B36 (2005) 307.
\item L. Cunqueiro, E.G. Ferreiro, F. del Moral and C. Pajares, hep-ph/0505197 (2005).
\end{enumerate}
\end{document}